\documentclass[aps,pre,nofootinbib,twocolumn]%showpacs,
{revtex4}
\usepackage{graphicx}
\usepackage{amssymb}
\usepackage{amsmath}
\usepackage{caption}
\usepackage{multirow}
\usepackage{array}
\usepackage{float}
\usepackage{subfigure}
\usepackage{color}
\usepackage{lipsum}

\begin{document}

\title {Effect Of Site Selective Ion Channel Blocking on Action Potential }

\author{Krishnendu Pal$^{1,2,+}$ and Gautam Gangopadhyay$^{2,}$}
\email{+pckp@iacs.res.in,*gautam@bose.res.in}

\affiliation{$^{1}$Indian Association for the Cultivation of Science, Jadavpur, Kolkata-700032, India.\\$^2$S N Bose National Centre for Basic Sciences, Block-JD, Sector-III, Salt Lake, Kolkata-700106, India.}

%\pacs{87.10.Mn, 87.16.Vy, 87.19.ll, 87.19.lb} 
\date{\today}
%\singlespace

\begin{abstract}
In this work we have theoretically investigated how the action potential generation and its associated intrinsic properties are affected in presence of ion channel blockers by adapting  Gillepie's stochastic simulation technique on a very basic neuron of Hodgkin-Huxley type. With a simple extension of the Hodgkin-Huxley Markov model we have mainly investigated three types of drug blocking mechanism such as (i) only  sodium channel blocking, (ii) only  potassium channel blocking and (iii) dual type blocking  and showed that the major experimental and physiological  observations such as  ionic currents, spiking frequency trends, change in action potential shape and duration, altered gating dynamics etc due to the presence of ion channel blockers can be well reproduced. Our results show that the nature of action potential termination process in presence of sodium and potassium channel blockers are  distinct and physiologically very different from each other. We have found that although the sodium and potassium ionic currents have interdependent relationship over a course of an action potential but sodium and potassium channel blockers have distinct signatures on ionic currents. In presence of only sodium channel  blockers the frequency of action potential generation falls off exponentially with increasing drug affinity, whereas in contrast, for only potassium channel blockers  initially an  enhanced spiking activity of action potential is found followed by a gradual decrease of the spiking frequency as the drug affinity increases. In case of dual type blockers with equal sodium and potassium channel binding affinity, the spiking frequency passes through maxima and minima due to the competition between channel number fluctuation and overall sodium and potassium conductances. We have found that sodium channel blockers shorten the  duration of action potential while the potassium channel blockers delay it which are of great physiological and pharmacological importance. We have also shown how the ion channel blockers alter the gating dynamics and such altered gating itself modulates the ion channel blocking which opens the possibility of  finding fundamental informations regarding probabilistic and dynamical features. Some experimental results  of  ion channel blocking in diverse systems  have been validated through our site selected binding scheme.
 Many other types of blocking mechanisms such as closed state blocking, inactive state blocking etc can also be explored using our method  with desired level of structural and functional details.   

\end{abstract}

\maketitle

\newpage
\section{Introduction}

Ion channels  are typically very complex transmembrane proteins\cite{Ashcroft} exhibiting a high degree of both structural and functional  diversity\cite{Hille}.  They have very distinct electrical potential dependent gating mechanism where the protein structures  can adopt several conformational states such as closed, open and inactive states in which states they can either transport ions across the membrane creating pores or restrict the ion permeability by closing the pore  when required\cite{Clare}. The ion channels are responsible for generating action potentials which are the basic requirement of cell to cell communication and signal propagation in nervous system\cite{Hille}. Among the various cation, anion and neutral ion channels  in general sodium and potassium channels are mainly responsible for action potential generation and its termination in most  of the neurons.  There exists certain compounds  or molecules, typically called as ion channel blockers which selectively bind to specific protein conformations of the ion channels and regulate their gating mechanism by blocking the passage of ions across the membrane. The high degree of specificity of these channel blockers on certain channels make them a valuable tool to treat numerous neural disorders\cite{Gilman, Gregorio, Catterall}.  Channel blockers of different types such as cationic blockers, anionic blockers, amino acids, and other chemicals regulate the functional properties of the channel or prevent them to respond normally. The naturally occurring sodium channel selective blocker TTX\cite{narahasi} was  known since 1964, followed by  Saxitoxin (STX), Neosaxitoxin (NSTX). Local anesthetics such as Lidocain, Phenytoin, Amiloride,  Bupivacaine and Tetracaine etc are clinically used\cite{Lipkind1, Tikhonov} for sodium channel blocking. On the other hand  potassium channel blockers, such as 4-aminopyridine and 3, 4-diaminopyridine etc are used as local anaesthetics\cite{Judge} and Tetraethylammonium (TEA) is used only for experimental purpose\cite{Catterall, Campbell}. A sodium blocker that blocks the open pore of the channel are called open state blocker\cite{Lipkind1, Scholz-Fozzard-kp, Wang}. Also there exists closed state blocker, inactive state stabilizing blockers etc\cite{Lipkind1, Scholz-Fozzard-kp, Wang}. Potassium channel blockers either bind with the selectivity filter  or at the  central cavity(open state) of the channel\cite{Piasta-Guo-Posson}.  From the discovery of TTX toxicity\cite{narahasi} the present knowledge of ion channel blockers have been a very vast, complicated and rigorous journey which was mostly developed for pharmacological interests\cite{Gilman, Gregorio, Catterall, Tikhonov, Campbell}. 

Recently Markov modelling and computer simulations prove to be promising techniques which provide new insights into the fundamental principle and mechanism about how these drugs alter the normal biological process\cite{Gregorio, Vladimir-Fink-rudy-Rudy1}. Single channel Markov models with discrete protein conformational states can simulate state specific channel properties and their alterations by mutations, disease or drug binding\cite{Vladimir-Fink-rudy-Rudy1}. As Markov models can be developed both at the level of single channel activity and at the macroscopic state, they provide an implicit relationship between the single channel recording and the macroscopic current\cite{Nekouzadeh-Nekouzadeh1}. Now the famous neuronal action potential model of Hodgkin and Huxley\cite{hodg} is based on an assumption that the ion permeation process can be approximated as both continuous and deterministic\cite{hodg} as it considers infinite number of ion channels inside a neuron. However, the permeation process existing within active membrane is now known to be neither continuous nor deterministic. Active membrane  consists of finite number of ion channels which undergo  random fluctuations between open and closed  states\cite{Hille} which scale inversely proportional to the number of channels present  in a particular patch of a neuron. Recent works also  reveal that fluctuation in the states of these channels are physiologically important in small neuronal structures\cite{DeFelice, Clay1983, Strassberg, Rubinstein}. When the number of ion channels inside a neuron cell membrane is finite  or small, the effect of internal noise become more and more important\cite{Jung,  DeFelice, Rubinstein, hangii2-hangii1-hangii4}. Channel number fluctuations can itself cause spontaneous spiking activity\cite{hangii2-hangii1-hangii4} without any stimulus and thereby the stochastic version of Hodgkin-Huxley model is proposed\cite{DeFelice,Strassberg,hangii2-hangii1-hangii4}. Therefore, the effect of channel noise can not be ignored also in the study of drug binding. Although innumerable number of experiments have been performed to study the channel blocking phenomena, the theoretical investigation and comprehensive understanding of the state specific blocking still requires detailed investigation from which the corresponding physiological consequences of drug blocking can be deduced that too from the level of action potential description itself. From the vast number of studies already made it is very difficult to comprehend the mechanistic link between the site specific blocking and the biophysical consequences of that. So  a microscopic reverse analysis is necessary from theoretical perspective  to validate the known observations.  

Inspired by the work done by Schmid, Goychuk and H\"anggi\cite{hangii3} on the effect of ion channel blockers using Langevin type of stochastic description, we present here a simple yet  detailed and realistic approach for channel blocking kinetics using the standard Markovian squid axon model of Hodgkin-Huxley\cite{hodg} which represents a basic neuron cell type.  We have incorporated new drug blocked states in the original Hodgkin-Huxley model to theoretically extend it to a drug binding model.  We then have suitably adapted Gillespie\cite{Gillespie} stochastic simulation  algorithm  to study the effect of ion channel blockers on action potential, ionic current, spiking frequency, action potential duration and gating dynamics in moderately stochastic limit of channel noise. We have proposed three types of drug blocking mechanisms, e.g,  sodium channel only blocking, potassium channel only blocking and a dual type blocking scheme with comparative binding affinity to these two channels. We have shown here through a few experimental evidences\cite{Storm, Brown1, Hlubek, bean1, bean2, Kullmann, Gullo, Brown, Saikawa, Colatsky, Nanasi, Belardinelli, mcgrawhill-perez} that our approach successfully correlates to the known physiological effects of ion channel blocking.

The Layout of the paper is as follows. In Section (II) we have discussed the kinetic scheme of drug blocking and the suitable modification of the simulation algorithm for the present purpose. Effect of channel noise on the action potential  has been discussed also under this section. In Section (III) the sodium and potassium channel blocking effects on the trains of action potential have been discussed with fundamental biophysical details. In Section (IV) comparative study on the effect of different types of blockers on ionic current, spiking frequency, action potential duration and gating dynamics has been discussed under various subsections. In Section (V) we have discussed how our results are in good agreement with the experimental and theoretical works done earlier. Finally the paper is concluded in Section (VI).

\section{Proposed Kinetic Scheme and  Stochastic Simulation of Action Potential In Presence of Channel Blockers}

%The famous neuronal action potential model of Hodgkin and Huxley\cite{hodg} is based on an assumption that the ion permeation process can be approximated as both continuous and deterministic\cite{hodg}. However, the permeation process existing within active membrane is known to be neither continuous nor deterministic. Active membrane  consists of discrete number of ion channels which undergo  random fluctuations between open and closed  states\cite{Hille}. Recent works also  reveal that fluctuation in the states of these devices may be physiologically important in small neuronal structures like nodes of Ranvier\cite{DeFelice, Clay1983, Strassberg, Rubinstein}.  At low patch size or small membrane area when less number of ion channels are considered the spontaneous firing can occur even without the presence of external stimulus(injected external current). The firing arises due to the stochastic fluctuation of the ion channels. These spontaneous firing rate decreases with  increase in the size of the membrane patch\cite{hangii2-hangii1-hangii4}. Thus the stochastic version of Hodgkin-Huxley model is proposed\cite{DeFelice,Strassberg,hangii2-hangii1-hangii4}.

The stochasticity in the Hodgkin-Huxley model is implemented in two ways such as theoretical approximation\cite{fox-fox-lu} method where it approximates the dynamics of the internal gating variables are governed by Langevin\cite{fox-fox-lu}  type of description of stochasticity, where the Gaussian white noise terms are added to the internal gating dynamical variables, described in the later part of this paper. The other one is the computational stochastic  simulation based on kinetic Monte-Carlo simulation of Hodgkin-Huxley Markov model\cite{DeFelice, Clay1983, Strassberg,  Gillespie, Skaugen,  Manwani, Koch, chow, white}. In this paper we have used the second one.

\subsection{Markov Model of Hodgkin-Huxley Action Potential Simulation}
Here we briefly describe  the Hodgkin-Huxley Markov model and its kinetic scheme. The Hodgkin and Huxley(1952)\cite{hodg} gating mechanism consists of four  activating $n$ gates  for potassium channel and its  Markov model consists of five states, such as, $n_0$(resting state, when all four gates are closed), \textbf{$n_1$}(only one gate is open), $n_2$, $n_3$ and $n_4$(open state or the single ion conducting state, when all four gates are open). The model of potassium channel thus have 8 transition rates designated by $\alpha_{n}$s and $\beta_{n}$s between these 5 states. On the other hand, the sodium channel has three activating $m$ gates with four distinct states and one inactivating $h$ gate with two distinct states. Thus the kinetic scheme based on Markov process has 8 states, such as, $m_0 h_0$(resting state), $m_1 h_0 , m_2 h_0 , m_3 h_0, m_0 h_1 , m_1 h_1 , m_2 h_1 , m_3 h_1$(open state or ion conducting state),  with a total of 20 transition rates designated by $\alpha_{m,h}$s and $\beta_{m,h}$s. Thus as a whole there are 28 transitions between 13 states to be considered for simulating action potential\cite{chow}. Each ion channel randomly fluctuates between these discrete states.
%\onecolumngrid
\begin{figure}[h!]
\centering
\includegraphics[trim={0.001cm 0 0 0},clip,width=9.0cm,angle=0]{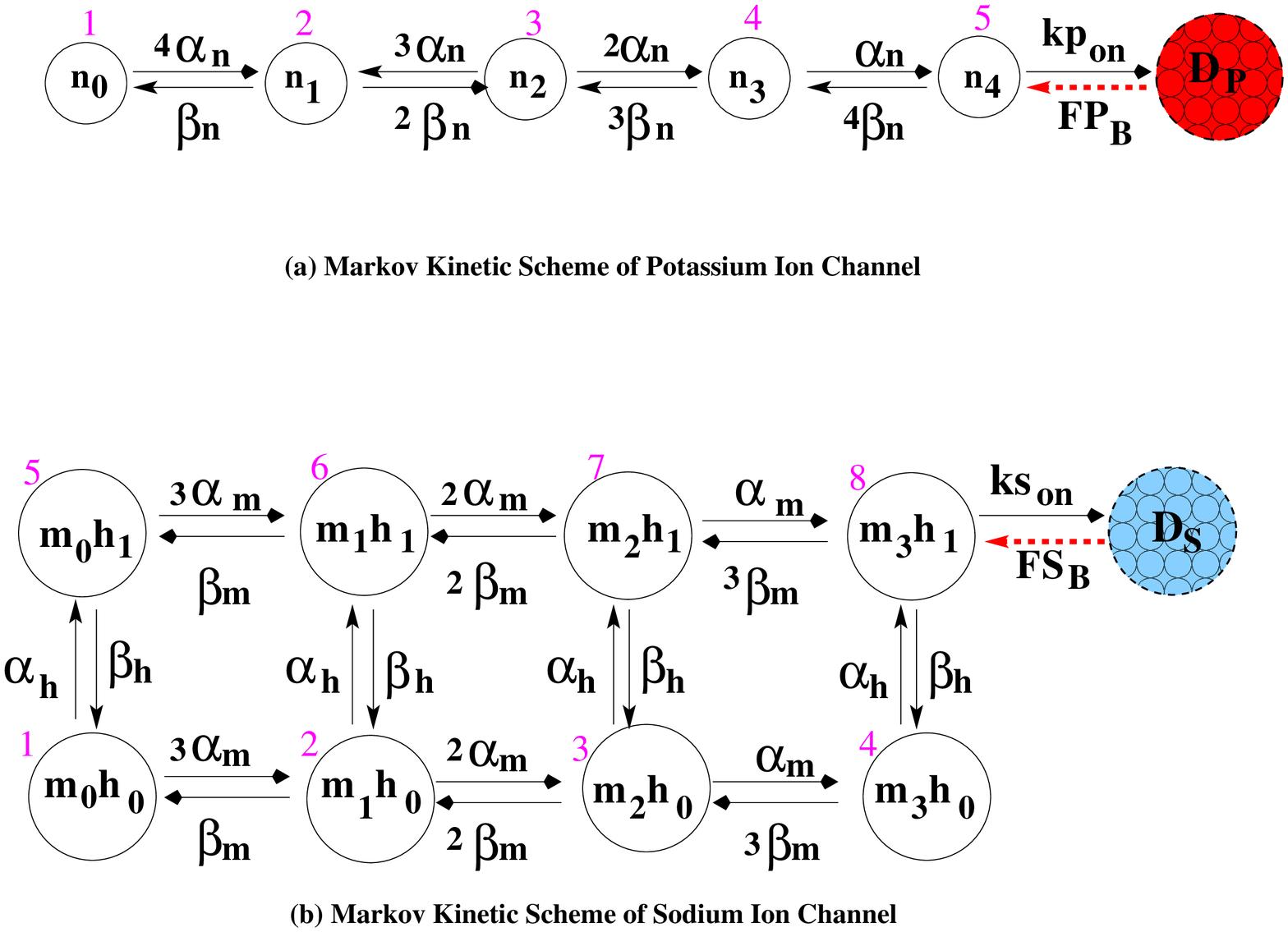}
\caption{\textbf{Markov model of Potassium and Sodium ion channel:} The potassium channel in figure (a) is a five state model where we have added an extra `drug-bound' state, $D_P$. Similarly in figure (b) sodium channel is an eight state model where we have added a `drug-bound' state, $D_S$.  }
\label{f1}
\end{figure}
%\twocolumngrid

The expressions of the voltage dependent rates\cite{hangii2-hangii1-hangii4} are given as follows $ \alpha_m(V)=(0.1(V+40))(1-\exp[-(V+40)/10])^{-1}$, $\beta_m(V)= 4\exp[-(V+65)/18]$, $\alpha_h(V)=0.07 \exp[-V+65)/20]$, $\beta_h(V)={1+\exp[-(V=35)/10]}^{-1}$, $ \alpha_n(V)=(0.01(V+55))(1-\exp[-(V+55)/10])^{-1}$, $\beta_n(V)=0.125\exp[-(V+65)/80].$  The Hodgkin-Huxley\cite{hodg} action potential or transmembrane voltage, V is given by the equation,
\begin{widetext}
\begin{equation}
C_m \frac{d}{dt} V(t)+ G_K(t)(V(t)-E_K)+G_{Na}(t)(V(t)-E_{Na})+G_{L}(V(t)-E_L)=I_{ext}(t).
\label{mem}
\end{equation}
\end{widetext}
Here V is in mV unit and the rates are in ms$^{-1}$.  Parameters  and their descriptions are given in TABLE \ref{t1} \cite{hangii2-hangii1-hangii4}. 
\begin{center}
\begin{table*}[t]
\begin{tabular}{|c|c|c|}
\hline
\hspace{0.5cm}$C_m$\hspace{0.5cm}  & \hspace{0.5cm} Membrane capacitance\hspace{0.5cm} & \hspace{0.5cm}1 $\mu$F/cm$^2$\hspace{0.5cm}  \\%\hline

\hspace{0.5cm}$E_K$\hspace{0.5cm}  & \hspace{0.5cm} K$^+$  reversal potential\hspace{0.5cm} & \hspace{0.5cm}-77.0 mV\hspace{0.5cm}  \\

\hspace{0.5cm}$\rho_{K}$\hspace{0.5cm}  & \hspace{0.5cm} K$^+$ channel density\hspace{0.5cm} & \hspace{0.5cm}18 channels/ $\mu$m$^2$\hspace{0.5cm}  \\

\hspace{0.5cm}$g_K^{max}$\hspace{0.5cm}  & \hspace{0.5cm} Maximal K$^+$ channel conductance(all K$^+$ channels are open)\hspace{0.5cm} & \hspace{0.5cm}36.0 mS/cm$^2$\hspace{0.5cm}  \\

\hspace{0.5cm}$\gamma_K$\hspace{0.5cm}  & \hspace{0.5cm} Single K$^+$  channel conductance\hspace{0.5cm} & \hspace{0.5cm}20 pS\hspace{0.5cm}  \\
\hline

\hspace{0.5cm}$E_{Na}$\hspace{0.5cm}  & \hspace{0.5cm} Na$^+$  reversal potential\hspace{0.5cm} & \hspace{0.5cm}50.0 mV\hspace{0.5cm}  \\

\hspace{0.5cm}$\rho_{Na}$\hspace{0.5cm}  & \hspace{0.5cm} Na$^+$  channle density\hspace{0.5cm} & \hspace{0.5cm}60 channels/$\mu$m$^2$ \hspace{0.5cm}  \\

\hspace{0.5cm}$g_{Na}^{max}$\hspace{0.5cm}  & \hspace{0.5cm} Maximal Na$^+$  channel conductance(all Na$^+$  channels are open)\hspace{0.5cm} & \hspace{0.5cm}120.0 mS/cm$^2$\hspace{0.5cm}  \\

\hspace{0.5cm}$\gamma_{Na}$\hspace{0.5cm}  & \hspace{0.5cm} Single Na$^+$  channel conductance\hspace{0.5cm} & \hspace{0.5cm}20 pS\hspace{0.5cm}  \\

\hline
\hspace{0.5cm}$E_L$\hspace{0.5cm}  & \hspace{0.5cm} Leak reversal potential\hspace{0.5cm} & \hspace{0.5cm}-54.4 mV\hspace{0.5cm}  \\

\hspace{0.5cm}$g_L$\hspace{0.5cm}  & \hspace{0.5cm} Leak conductance\hspace{0.5cm} & \hspace{0.5cm}0.3 mS/cm$^2$\hspace{0.5cm}  \\

\hline
\end{tabular}
\caption{Parameters of Hodgkin-Huxley equation\cite{hangii2-hangii1-hangii4}. } 
\label{t1}
\end{table*}
\end{center}
For a discrete stochastic channel populations  the potassium and sodium membrane conductances across are expressed by the following equations (\ref{cond}),
%\begin{widetext}
%\begin{equation}
\begin{subequations}
\label{cond}
\begin{eqnarray}
G_K(V,t) & =&  g_K^{max} \left[N_{n_4}/N_{K}\right],\label{equationa}
\\
G_{Na}(V,t)& = &  g_{Na}^{max} \left[N_{m_3h_1}/N_{Na}\right],\label{equationb}
\end{eqnarray}
\end{subequations}
%\end{equation}
%\end{widetext}
where $N_{n_4}$ and $N_{m_3h_1}$ are the number of potassium and sodium channels in  \textbf{open} state, respectively and $N_{K}$ and $N_{Na}$ are the total number of potassium and sodium channels present in the membrane patch considered, respectively. Solving the stochastic simulation one obtains the population of individual states from which the open state populations for sodium and potassium channel are put into conductance equation (\ref{cond}) and then solving equation (\ref{mem}) one obtains the action potential.

\subsection{Proposed Scheme of Channel Blocking}
	Now, in this work we have studied the effect of ion channel blockers on action potential. Ion channel blockers can bind to several conformational states of ion channels such as closed state, inactivated state and open-state of ion channels\cite{Hille}. However, here we have only considered the case of those blockers which preferentially bind to the open states of the channels.    As these blockers block the open-pore of the channel and hinder ion permeation,  we can think of a state, say,``drug bound" state which is  accessible only if the channels  enter the open state. Thus the drug bound state is coupled to the open state of the sodium or potassium channel. The drug bound state basically represents the state of channels which are blocked by the blockers. To implement this on potassium channel we designate this drug bound state as $D_P$ and for sodium channel it is $D_S$ as seen from figure (\ref{f1}). The forward transition rate for sodium or potassium blocking are given by `k{x}$_{on}$', where x=s for sodium blockers and x=p for potassium blockers. The affinity of drug binding is usually defined as kx$_{on}$= kx$_b$*[D], where kx$_b$ is the binding constant of the drug and [D] is the concentration of the drug\cite{Scholz-Fozzard-kp}. Thus any change of the value of `k{x}$_{on}$' means the change of concentration of the blockers or drugs of a particular type having specific binding constant.
	
Now the backward transition for reopening is a very  slow process compared to the forward. Channel blockers or local anesthetics take around 1-4 hours(duration of action) before they are completely removed\footnote{Amide blockers(lidocaine, bupivacaine, mepivacaine etc) are biotransformed in the liver, ester blockers(cocaine, benzocaine, procaine etc ) are hydrolyzed in the bloodstream by plasma esterases or the hydrolysis  of the side chains of the blockers make them inactive and then they are eliminated via blood circulation\cite{la1-la2-la3-la4}.} from the channel proteins\cite{la1-la2-la3-la4}. In general, clinically used channel blockers or local anesthetics can be divided into three categories: short acting (e.g., 2-chloroprocaine, 45-90 minutes), intermediate duration (e.g., lidocaine, mepivacaine, 90-180 minutes), and long acting (e.g., bupivacaine, levobupivacaine, ropivacaine, 4-18 hours)\cite{la1-la2-la3-la4}. Thus it is seen that elimination of the drug from the site or the recovery of the channel is a very slow process compared to that of drug binding where the onset of local anaesthetic actions take place within few minutes(\textit{as we shall also see that the simulation runs with the transition rates mentioned earlier take few seconds only to completely terminate the action potential generation for a patch area we have considered in this paper}).  Thus the backward flux can therefore reasonably be considered as a very small valued constant during the course of drug binding action. We designate the backward flux as Fx$_B$=0.001, a constant throughout\footnote{There is roughly 18,000  channels in total are present in 200 $\mu m^2$ patch area(area that we have chosen in this work). If we even consider the blocking action prolongs for 1 hour, then the approximate flux becomes Fx$_B \approx \frac{1}{1 \times 3600 \times10^3}\times 18000 \approx 0.005 $. Now in fact not exactly 18000 blocked channels will be there. Blood circulation hinders the channel blockers to penetrate the lipid bilayers by washing them out and a certain fraction of the blockers can enter to block channels, thus have lesser chance to block all the channels, also density of the patch is not homogeneous everywhere that some patches can have lesser number of channels. Besides some drug actions last more than 4 hours. Considering these facts together we have taken a lower value which is 0.001(easy to compare), which hardly impacts the nature of the results as we have verified.}.  Now, we have gradually increased the drug binding affinity, kx$_{on}$ from 0.0001 to 1.0 or more, keeping Fx$_{B}$=0.001, constant. When we consider only sodium blockers we do not consider the potassium blocking state, $D_P$ and vice verse except for dual type of blocking as we shall discuss later.  Thus with the attached single drug-bound state  Hodgkin-Huxley model now have 14 states and 30 transition rates. 

\subsection{Stochastic Ion Channel Simulation Using Gillespie Algorithm}

The computer based stochastic simulation algorithm using Markovian model  can be classified into two algorithms, such as, i) Channel-State-Tracking(CST) algorithm and ii) Channel-Number-Tracking(CNT) algorithm. The CST algorithm tracks the \textit{specific states of each channel} and superimposes individual channel currents corresponding to the states. This algorithm although simple but it is computationally very costly and intensive\cite{Clay1983, Strassberg, Rubinstein}. However an easy alternate is CST algorithm using Gillespie method\cite{Gillespie, Skaugen,  Koch, chow}  which  keeps track of the \textit{number of channels in each state} with the  assumption that multi-channel systems are independent and memoryless. This Channel Number Tracking(CNT) algorithm provides much greater efficiency in cases where many channels are simulated as the algorithm calculates an effective transition rate associated with the multi-channel system by allowing only one transition among all states in a random time interval. Thus in this paper we have utilized Gillespie's CST simulation algorithm to study the effect of ion channel blockers. The simulation method is very well known in the literature\cite{Gillespie,Skaugen,Koch,chow}. The steps that we considered to develop the stochastic algorithm for the site selective binding of sodium and potassium blockers for a particular patch size of a neuron have been elaborately discussed in \textbf{Appendix (A1)}.

\subsection{Effect of Channel Number Fluctuation}

Next we have studied the action potential and conductance of sodium and potassium channels without the presence of external stimulus or external current, $I_{ext}=0.0 \mu A/cm^2$. We  start from very high stochastic limit of channel noise with patch size as low as $A=1.0 $ $\mu m^2$ and extend it to the deterministic limit such as $A=200$ $\mu m^2$. Here we do not consider the presence of any drug, hence kx$_{on}$ = Fx$_{B}= 0.0$. Thus here we  consider 13 states and 28 reactions: the original Hodgkin-Huxley model.
%\vskip 0.2cm
\begin{figure}[h]
\centering
\includegraphics[trim={0.001cm 0 0 0},clip,width=9.0cm,angle=0]{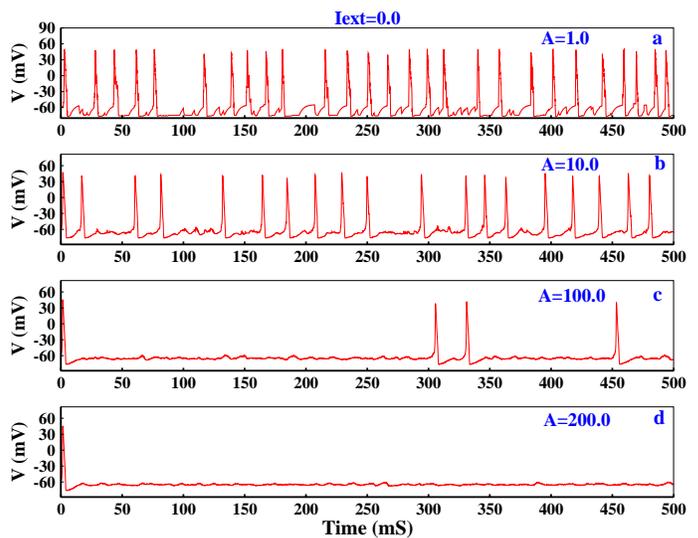}
\caption{\textbf{Effect of channel number fluctuation on action potential:} Action potentials of various patch size are plotted here. From (a) to  (d) the action potentials are shown for A= 1.0, 10.0, 100.0, 200.0 $\mu m^2$ at $I_{ext}=0.0$ $ \mu A/cm^2$. It shows how the effect of channel noise and consequent spontaneous action potential generation falls off as the deterministic limit is approached. }
\label{f2}
\end{figure}
In figure (\ref{f2}) we have shown how the action potential vary from patch size, A=1.0 to 200 $\mu m^2$. As seen from figure \ref{f2}(a) that at very low patch size  channel noise plays very important role. These channel number fluctuations can alone originate spontaneous spiking activity\cite{hangii2-hangii1-hangii4} without even the presence of external current. As soon as the patch size is increased the spontaneous spiking rate decreases. As we can see at moderately high patch size, for example,  $A=200 \mu m^2$, the spiking phenomenon vanishes away and system behaves similarly as it does in deterministic limit at  $I_{ext}=0.0 \mu A/cm^2$. 

As a side note we would like to mention here that the Gillespie method that we have used to simulate action potential, although is a very popular method and often claimed to be an exact simulation technique in literature, however, contrary to that popular belief it can not be claimed as an exact method especially while considering very low patch sizes. It is basically an approximate method which works fine for higher patch sizes. A brief discussion about its applicability has been discussed in \textbf{Appendix (A2)}. Exact simulation for figure \ref{f2} (a) and (b) is beyond the scope here. These two plots are not exact but here their sole purpose is to convey the message that at low patch sizes the channel fluctuations do play very important role.

%\begin{figure}[h!]
%\centering
%\includegraphics[trim={0.001cm 0 0 0},clip,width=10.0cm,angle=0]{r3.eps}
%\includegraphics[width=10.0 cm,keepaspectratio]{r3.eps}
%\caption{Conductance curves for sodium and potassium channels from microscopic regime to deterministic limit. In the left traces potassium and in the right traces  sodium channel conductances are plotted for $A=0.5, 1.0, 10.0, 200$ $\mu m^2$ at  $I_{ext}=0.0 \mu A/cm^2$.} 
%\label{r3}
%\end{figure}
%The stochastic nature of conductance for both sodium and potassium channels are well observed from the figure (\ref{r3}). At high patch size the conductance curves behave similar to the response observed in the deterministic limit.

%\newpage
\section{Sodium and Potassium Channel Blockers}
\label{sec3}
Now we come to the main focus of this work, i.e. studying stochastic drug binding kinetics in ion channels. To see the effect on action potentials we have gradually increased the external stimulus from zero to higher values and seen that at $I_{ext}= 6.9$ $\mu A/cm^2$ they show considerable amount of spiking activity. The values of ionic current lesser than 6.9 $\mu A/cm^2$ show very irregular spiking pattern with higher intervals which requires very long time simulations and become very  clumsy to be presented in plots. At this selected value of ionic current they show considerable amount of spiking activity with lesser intervals which can be plotted very nicely and helps us to interpret the effect of drugs on action potential spiking trend. However the results in general discussed here show exactly similar behavior for all other choices of  $I_{ext}$ other than zero. We have kept the patch size,  A=200 $\mu m^2$ for which the effect of channel number fluctuation is moderately close to the deterministic limit but not exactly the deterministic limit, because with the value of $I_{ext}=6.9$ $\mu A/cm^2$, a considerably higher patch size is required to match deterministic result. On the other hand at very low patch size the channel noise makes it almost impossible to study the binding effects. At this value of patch size, A=200 $\mu m^2$, we are dealing with almost 12000 sodium channels and 3600 potassium channels.

\subsection{Only Sodium Channels Are Blocked}
As mentioned earlier, here we have considered only the open state sodium channel blockers in this  study. Here we do not add any potassium blockers together. Thus here only $D_S$ state is present not $D_P$.  
\begin{figure}[h]
\centering
\includegraphics[trim={0.001cm 0 0 0},clip,width=9cm,angle=0]{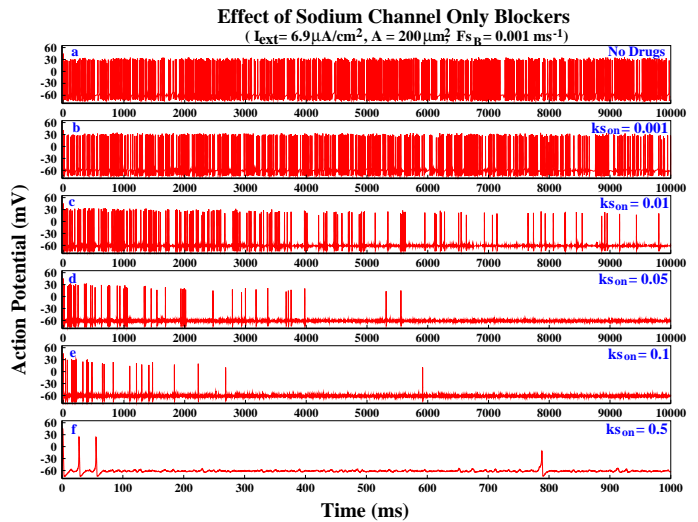}
\caption{\textbf{Effect of sodium channel only blockers:} Action potentials in presence of sodium channel only blockers with increasing affinities are plotted here. In figure (a) the action potentials are plotted without the presence of any drug. From figure (b) to (f) the action potentials are plotted for ks$_{on}$=0.001, 0.01, 0.05, 0.1 and 0.5 with A= 200 $\mu m^2$, $I_{ext}=6.9$ $\mu A/cm^2$ and reopening flux, $FS_B$= 0.001 ms$^{-1}$. In all the cases where Na-blocker is present the action potential train ultimately dies off, sooner for higher affinity(not shown for (b) and (c) as they require more than 20,000 ms, plotting them makes the plots compact and clumsy, the train dies off similarly as shown in (d)-(f)). } 
\label{f3}
\end{figure}
In figure (\ref{f3}) we have shown the effect of sodium channel blockers on action potentials train with time with increasing drug affinity or concentration. It is seen that in presence of drug the spiking activity gradually dies off temporally. With increasing affinity the spiking activity dies off in a faster rate. As seen from figure \ref{f3}(f) at very high value of $ks_{on} = 0.5$ the system almost fails to regenerate action potentials. As the affinity increases more channels quickly go to the drug-bound state and gets trapped in the drug bound state. Thus the available  open sodium channels gradually decreases which makes it difficult for that particular patch to generate action potential. Beyond $ks_{on}= 1.0$ we have seen that the patch totally fails to generate even a single spike after the first spike.

\subsection{Only Potassium Channels Are Blocked}
Next we have studied the effect of open state potassium channel blockers on the action potential generation. Here the $	D_P$ state is only considered, not $D_S$ state. 
\begin{figure}[h!]
\centering
\includegraphics[trim={0.001cm 0 0 0},clip,width=9cm,angle=0]{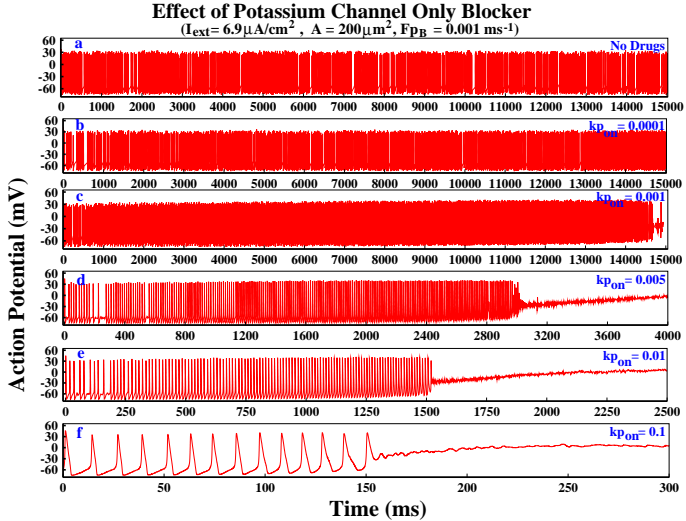}
\caption{\textbf{Effect of potassium channel only blockers:} Action potentials in various blocking affinities of potassium only blockers are plotted here. In figure (a) the action potentials are plotted without the presence of any drug. From figure (b) to (f) the action potentials are plotted for kp$_{on}$ = 0.0001, 0.001, 0.005, 0.01 and 0.1 for A = 200 $\mu m^2$, $I_{ext}=6.9$ $\mu A/cm^2$ and reopening flux, $FP_B$ = 0.001 ms$^{-1}$. Although here also action potential dies off temporally but the nature of termination is quite different from that of the sodium channel only blocker. } 
\label{f4}
\end{figure}
In figure (\ref{f4}) the action potentials for various kp$_{on}$ are plotted. Here also for all the cases where drug is present ultimately the spiking activity of the action potential  dies off temporally. As usual with increasing affinity the blocking of action potential occurs in a faster rate. But, surprisingly here we have observed an interesting trend of action potential termination which is quite different from sodium blocking case. 

If we carefully notice the spikes in figures (\ref{f3}) and (\ref{f4}) at high affinity regions, we can see that in presence of sodium channel blockers the number of generated action potentials although decreases down significantly, but every spike has a complete shape(they depolarize $>$ repolarize $>$ hyperpolarize $>$ again comes back to resting potential). But for potassium blocking case we see that after a few millisecond the spiking activity entirely falls off with the last spike having incomplete shape. This happens due to lack of available open potassium channels.  Due to the lack of available open potassium channels the repolarization process gets hampered, leading to the incomplete generation of action potential. Once it is not complete further  generation of action potential is not possible as the refactorization process which was necessary for further generation of action potential could not complete. Thus no action potential generation can take place further as seen from figures \ref{f4}(d-f). But in case of sodium channels there remains plenty of available potassium channels which can restore the depolarized potential back to the resting one. Thus in sodium blocking case the spikes generate with complete shape even if there number decreases with increasing affinity.

%\newpage
\section{ Effect Of The Channel Blockers}
In this section we have shown how the sodium and potassium channel blockers and  local anesthetics(dual type blockers) affect  the ionic current, spiking frequency, duration of the action potential and gating dynamics, all of  which plays physiologically very significant roles.  

\subsection{Ionic Current}   
The sodium and potassium ionic currents across the membrane are given by the eqution (\ref{ion-curr}). %The sodium  current is negative thus it has inward direction of flow  and the potassium current is positive having out of direction of flow.   Now here we would like to point out that the sodium and potassium currents have interdependent relationship on each other over a course of  an action potential. 
Now here we would like to point out that although sodium and potassium channels  are two different proteins,  during an action potential, they work together and thus sodium and potassium currents have interdeppendent relationship. The number of spikes of sodium ionic current is equal to the number of spikes of potassium ionic  current. Hampering sodium channel conductance with sodium channel blockers  may not directly  affect the gating mechanism inside the  potassium channel proteins, but the decreased efficiency of conductance leads to decrease in the number of sodium ions coming inside and thus less number of potassium ions are required go out side to balance that depolarization, leading to a decrease in the magnitude of potassium current too.   On the other hand if we keep on decreasing the ionic conductance of potassium channels by applying potassium blockers, the repolarization process  or refactorization process would gradually get hampered which in turn will also affect the membrane potential which  eventually   becomes unsuitable for sodium channels to open as we have seen from figure (\ref{f4}). Thus both of the currents are affected if anyone of them is disturbed. 

\begin{figure}[h!]
\centering
\includegraphics[trim={0.001cm 0 0 0},clip,width=9.0cm,angle=0]{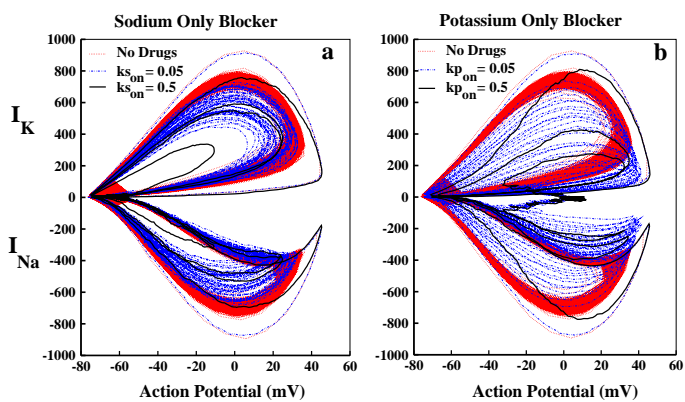}
\caption{\textbf{Ionic current:} In this figure we have plotted the sodium and  potassium channel ionic currents together with their corresponding  action potentials. In the left panel, figure (a), the effect of sodium channel only blockers with affinity, $ks_{on}$= 0.05, 0.5 are compared with the case where no drug is present. The top trace represents the potassium ionic current and the bottom trace represents the corresponding sodium ionic current. In the right panel similar plot has been shown with potassium only blockers. A= 200 $\mu m^2$ and $I_{ext}=6.9$ $\mu A/cm^2$. Both the blockers have different signatures of blocking. } 
\label{f5}
\end{figure}

Now in figure (\ref{f5}) we have shown the interdependent nature of sodium and potassium currents and how the currents are being affected in presence of ion channel blockers with increasing drug binding affinity. The currents are generated in response to a change in the respective ion channel conductances. The total number of Na$^+$ or K$^+$ ions that permeate inside or outside the membrane during the action potential is proportional to the areas under the individual current curves in the figures.

The red dotted curves in each traces of figure \ref{f5}(a) represents the case of drug free situation.  After the first spike the loop area shrinks. Afterwards the area of ionic currents slightly change due to the presence of the fluctuation in the number of available open channels which creates the broadening of the band. The blue dot-dashed curve for affinity $ks_{on}$=0.05 and and black-solid line for $ks_{on}$= 0.5 shows that both of the ionic currents gradually shrinks over time. If we plot in a 3D diagram with  membrane depolarization in x axis,  currents in y axis and time in z axis we see that with increasing time the area of the ionic currents shrink gradually with corresponding decrease in the peak height of the action potentials. This shows the gradual decrease of sodium ion influx inside the cell membrane.  

Next we have observed  an interesting feature of potassium channel  blocking on both the ionic currents. Unlike the case of sodium  blockers in  \ref{f5}(a), here we see that the loop area of both the ionic currents falls down very rapidly covering the entire space which indicates the decrease in the potassium channel conductance over time. Unlike the sodium only blockers, the corresponding action potential magnitude is much higher here when the currents fall down to zero. This once again clearly says that the membrane repolarization process is drastically hampered. As the loop areas are proportional to the number of ions being permitted across the membrane, the potassium only blockers show this amount is hampered more rapidly than sodium blockers of equal affinity.  The effect of blockers on effective ionic current, $I_{int}=I_{Na}+I_K+I_{L}$ over the corresponding action potentials can also be found in  the supplemental materiel figure (\ref{S1}).

\subsection{ Spiking Frequency}

\begin{figure}[h!]
\centering
\includegraphics[trim={0.001cm 0 0 0},clip,width=9.0cm,angle=0]{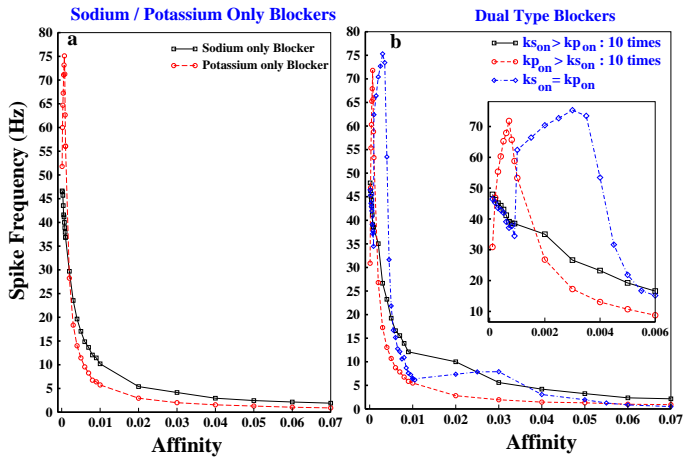}
\caption{\textbf{Spiking frequency:} In figure (a) the spiking frequency(Hz) of the action potentials in presence of increasing sodium and potassium channel only blocking affinity, ks$_{on}$ and kp$_{on}$ have been plotted respectively, ranging from 0.0001 to 0.07.   The spiking frequency is calculated over a long simulation run upto 20s. The squared-(black)-solid line and the diamond-(red)-dashed line shows the frequency of spiking of action potentials in presence of only sodium  and potassium channel blockers, respectively. In figure (b) the spiking frequency profile is plotted for the dual type blockers for the cases: (i) $ks_{on}>kp_{on}$: squared-(black)-solid  line, (ii) $kp_{on}>ks_{on}$: circled-(red)-dashed line and for (iii) $ks_{on}=kp_{on}$: diamond-(blue)-dot dashed line. For all the cases  A = 200 $\mu m^2$ and $I_{ext}=6.9 $ $\mu A/cm^2$ and those spikes are only considered which have peak heights more than -10 mV.  } 
\label{f6}
\end{figure}

Next we have studied  the effect of the two types of blockers on the spiking frequency of the action potentials with increasing kx$_{on}$ in figure \ref{f6}(a). 
Here we have found an interesting difference of spiking frequency trends between sodium and potassium blockers. It is seen that in presence of sodium channel blockers the spiking rate falls off exponentially with increasing affinity. But in presence of potassium channel blockers the spiking frequency initially with increasing kp$_{on}$ increases unlike sodium channel and then it decreases. This initial increase in spiking activity in presence of potassium blockers is consistent with the earlier works done in literature \cite{hangii3, Kocsis}. This happens because with the increase in affinity number of potassium channels decrease which causes an increase in the internal  noise and therefore channel number fluctuation starts playing significant role by spontaneously generating action potentials, as we have already seen from the figure(\ref{f2}). Actually in presence of potassium blockers there exists a competition between the overall conductance of the ions and the channel noise.  In the low affinity region the influence of the channel noise dominates over the ion conduction and thus the spiking activity increases. After a certain kp$_{on}$ value, the spiking activity decreases as the decreased overall conductance dominates here. For sodium channel only blockers, the over all conductance always dominates over channel noise. 

%\newpage
\subsection{Action Potential Duration}

Next we have seen the effect of channel blockers on the action potential duration(APD) itself. APD varies from one neuron to another neuron type. Any alteration of APD of a particular neuron can lead to significant complexities in signal transmission process and can give rise to critical physiological disorders. We have found that in presence of blockers APD significantly alters with time. First we have chosen three drug affinity regions such as  low, medium and high, corresponding to the sodium and potassium blocking. For sodium blockers we have chosen the affinities, ks$_{on}$= 0.005(low), 0.05 (medium) and   0.1(high). For potassium blockers we have chosen the affinities, kp$_{on}$ = 0.005(low), 0.01 (medium) and   0.05(high). The regions are different because with the same magnitude of affinity potassium blockers affect the spiking activity more rapidly than sodium blockers as seen comparing figures (\ref{f3}) and (\ref{f4}). Next for each affinity we have arbitrarily chosen a spike from few initial spikes and then compared it to the randomly selected spikes from intermediate and end positions of the spike trains and plotted them in a same scale.

\begin{figure}[h!]
\centering
\includegraphics[trim={0.001cm 0 0 0},clip,width=9.0cm,angle=0]{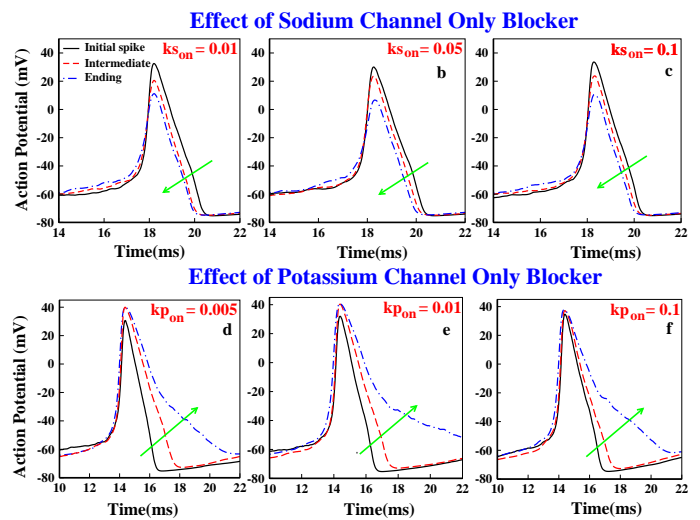}
\caption{\textbf{Action potential duration:} The temporal change of APD in presence of ion channel blockers have been shown here. In the top traces, (a-c), the initial, intermediate and end point spikes are plotted in a same scale for sodium channel only blockers with affinity ks$_{on}$ = 0.005(low), 0.05(medium) and  0.1(high ). In bottom traces, (d-f) similar plots have been done for potassium channel only blockers with affinities, kp$_{on}$ = 0.005(low), 0.01 (medium) and   0.05(high). In both cases, A= 200 $\mu m^2$ and $I_{ext}=6.9$ $\mu A/cm^2$. The green-solid arrows are indicating the direction of shortening or broadening of the incoming action potential. } 
\label{f7}
\end{figure}

From figure (\ref{f7}) we have again found an interesting difference in APD for two types of blockers. It is seen that for sodium channel only blockers  with increasing time the repolarization  process gets faster or the shortening of the APD occurs where as contrastingly, for potassium blockers the repolarization phase is seen to be delayed considerably with increasing  APD.  The shortening of APD in presence of sodium blocker happens because the effective  number of sodium ions are entering the cell decreases with time and thus less number of potassium ions required to go out to bring back the repolarization which thus gradually takes lesser time than usual. On the other hand in presence of potassium blockers the repolarization phase is seen to be delayed temporally because the number of available open potassium channels gradually decrease and thus the repolarization brought by the decreased number of channels take longer time. Thus the broadening of APD occurs gradually. This shortening or the broadening  of the APD is actually a special characteristic feature of sodium and potassium channel blockers respectively which have great physiological significance and clinical use. Patients with longer or shorter APD due to mutagenic, hereditary or other physiological conditions are treated with ion channel blockers.

\subsubsection*{Dual Blockers}

Here we want to point out that some  ion channel blockers such as local anesthetics are well known for their non-specific blocking nature. Also at an elevated concentration some drugs that primarily target Na channel may also seen to affect K channels. Channel specific type of blocking as discussed so far is more pronounced for TTX(specifically binds to Na channel only) and TEA(binds to potassium channel only) types of blockers only which are mostly used in experimental researches. But for local anesthetics or clinically used drugs we must consider both type of blocking simultaneously. For that purpose we have chosen three types of blocking schemes which summarizes all possible kinds of nonspecific binding mechanisms. For the first case we have kept both the sodium($ks_{on}$) and potassium($kp_{on}$) drug binding affinities equal, i.e, $ks_{on}=kp_{on}$. So these represents those class of drugs which binds to both type of channels with equal affinities. For the second case we have chosen those types of drugs which have approximately ~10 times higher affinity of blocking sodium channels\cite{Scholz-Fozzard-kp}. The major difference between local anesthetic action in  K$^+$ currents compared with Na$^+$ currents is the  lower affinity(approximately 10 times lower) in the former\cite{Scholz-Fozzard-kp}. So here we have kept the sodium blocking affinity, $ks_{on}>kp_{on} \approx 10$ times. Finally we kept $kp_{on}>ks_{on} \approx 10$ times. One can change the ratio according to the need or according to the knowledge of the binding affinity of a particular drug. %We have already tried many combinations but to show distinctly observable effect on action potential duration and spike frequencies we have limited the   $ks_{on}/kp_{on}$ or $kp_{on}/ks_{on}$ ratio to 10. 
The figures of action potential trains for these three categories of blocking can be found in the supplemental figures (\ref{S2}),  (\ref{S3}) and (\ref{S4}).

\textbf{Spiking Frequency:} Now we have plotted the spiking frequency profile for these three cases in figure \ref{f6}(b).  For the case of $ks_{on}=kp_{on}$(diamond-blue-dot-dashed line) it is seen  that initially frequency decreases  up to around ks$_{on}$=kp$_{on}$=0.001 and then passing through a minimum  it increases between a small region upto around 0.004 and then a large decrease is observed until 0.01 is reached. After that again an  increase in frequency is observed until 0.03 and then it falls off gradually. These sort of double minimum and maximum in frequency of spiking actually indicates a competition between overall sodium and potassium ionic and channel noise activity. The regions where the maxima  exist are  basically the regions where the channel number induced spontaneous spiking activity dominates on overall ion conductance\cite{hangii3}. In other words it can be said that the maximums arise due to the coherence resonance due to channel noise\cite{hangii2-hangii1-hangii4}. For the other two cases such as  $ks_{on}/kp_{on}\approx 10$(squared-black-solid line) and $kp_{on}/ks_{on}\approx 10$(circled-red-dashed line), the previous trends of sodium and potassium only blockers are observed as seen from figure \ref{f6}(a). The case of  $ks_{on}=kp_{on}$ is different which indeed reduces to the other two cases when the ratios become 10.  However, for all the three cases it is seen that the action potential generation gradually ceases off with an the unsuitable refactorization potential which clearly indicates that towards the end of the action potential train, potassium channel blocking plays more influential role in the action potential termination process.    

\begin{figure}[h!]
\centering
\includegraphics[trim={0.001cm 0 0 0},clip,width=9.0cm,angle=0]{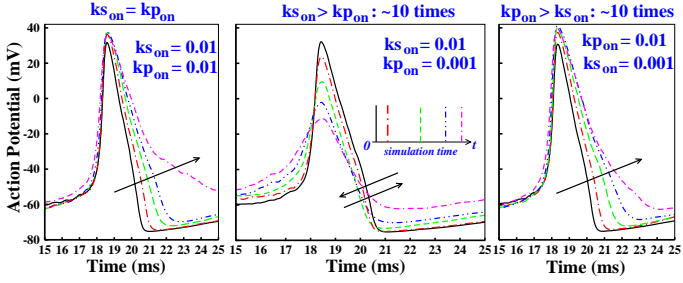}
\caption{\textbf{Action potential duration affected by dual blockers:} The change in the duration of the action potential in presence of local anesthetic channel blockers at a higher affinity  over their individual course of action potntial train have been plotted for the cases $ks_{on}=kp_{on}$(left),  $ks_{on}>kp_{on}$(middle) and  $kp_{on}>ks_{on}$(right).  The approximate positions of the five randomly chosen spikes over a train of action potential simulation run of t ms are shown in the inset of middle figure. Here also  A= 200 $\mu m^2$ and $I_{ext}=6.9$ $\mu A/cm^2$.} 
\label{f8}
\end{figure}

\textbf{Action Potential Duration:} Next we have plotted the effect of these three types of drug blocking mechanisms on the duration of the action potential in figure (\ref{f8}) over one individual action potential train. For the ks$_{on}$ $>$ kp$_{on}$ case(the middle trace), we have found that initially the APD shortening occurs for few ms due to the dominant(ks$_{on}$ 10 times stronger than kp$_{on}$) blocking of sodium channels over potassium channels. But shortly after  when sufficient amount of potassium channels are blocked, the effect of potassium blocking starts playing significant role by delaying APD. The inward arrow in the middle trace of figure (\ref{f8}) shows the shortening of APD followed by an outward arrow indicating the gradual delay of APD as time progresses. We have verified this for all affinity region. For higher affinity region, such as ks$_{on}$ = 0.1, the broadening of APD occurs, preceded by a quick APD shortening. However, here we want to point out that the shape of action potentials in this case have been found to be heavily affected such as the gradual decrease of the peak height of action potentials as time progresses(see supplemental  figure S3). Keeping the peaks fixed we have found that the initial upward depolarization phase of action potentials also changes significantly, as seen in the left portion of the peak in the middle trace. So the change of APD here, particularly in this case is considered  as the peak after shortening or broadening. In the right trace, the case $kp_{on}>ks_{on}$ is shown which shows exactly similar nature of APD broadening as shown in potassium channel only blocking case in figures \ref{f7}(d-f). For the $ks_{on}=kp_{on}$ case in left trace, although the frequency response curve came up with different signature but for the duration  of the action potential it basically displays the potassium blocking dominance by delaying action potential duration. Following Table(\ref{t2}) summarizes our results:

\begin{center}
\begin{table*}[t]
\begin{tabular}{|c|c|c|}
\hline
\textbf{Blocking mechanism}  & \textbf{Spiking Frequency} & \textbf{APD}  \\
\hline
Sodium channel only blocking  & Exponential decrease & Shortening occurs \\
\hline
Potassium channel only blocking & Initially increase and then decrese & Broadening occurs\\
\hline
Equal affinity :$ks_{on}=kp_{on}$ & Maxima and minima arises & Broadening occurs(potassium blocking dominates)\\
\hline
Preferential sodium blocking :$ks_{on}>kp_{on}$ & Exponential decrease & APD shortens initially, then broadening occurs\\
\hline
Preferential potassium blocking :$ks_{on}<kp_{on}$ & Initially increase and then decrese & Broadening occurs\\
\hline
\end{tabular}
\caption{Various blocking mechanism and their effect on spiking frequency and action potential duration(APD). } 
\label{t2}
\end{table*}
\end{center}
\subsection{ Gating Dynamics}

\begin{figure}[h!]
\centering
\includegraphics[trim={0.001cm 0 0 0},clip,width=9.0cm,angle=0]{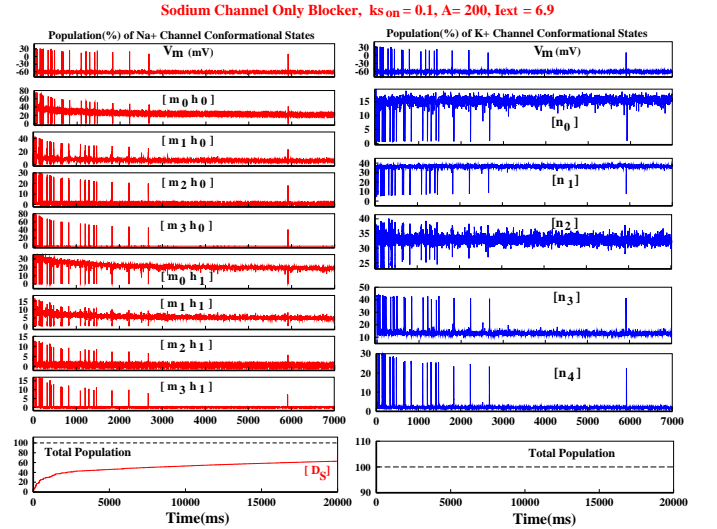}
\caption{\textbf{Gating dynamics in presence of sodium only blocker:} The action potential and the corresponding population of every sodium(left traces) and potassium(right traces)  channel conformational states are plotted for sodium channel only blocker with ks$_{on}$= 0.1 at A= 200.0 $\mu m^2$ and $I_{ext}$= 6.9 $\mu A/cm^2$.} 
\label{f9}
\end{figure}

Here we have studied the percentage of the total population of channels present at a particular time in different conformational states including the drug bound state. In the left traces of figure (\ref{f9}) the population(\%) for sodium only blockers for affinity, ks$_{on}$= 0.1 and the corresponding population(\%) of potassium channel states  has been shown in  right traces. Observing both the traces we find that each of the action potential spike is associated with either upward or downward spike in all the population plots,  both for sodium and potassium conformational states. This indicates that sudden random action potential generation is associated with sudden dramatic changes of the entire population occupancy of  all the conformational states.  From the left trace of figure (\ref{f9}) we see that in 20 seconds( although spikes occurring ceases long ago) only $\sim$50\% channels are trapped in drug bound state, $D_S$ which slowly increases with time. The rest of the population is mainly trapped in the inactive states and closed states. With increasing affinity the drug bound state populates in a faster rate.

\begin{figure}[h!]
\centering
\includegraphics[trim={0.001cm 0 0 0},clip,width=9.0cm,angle=0]{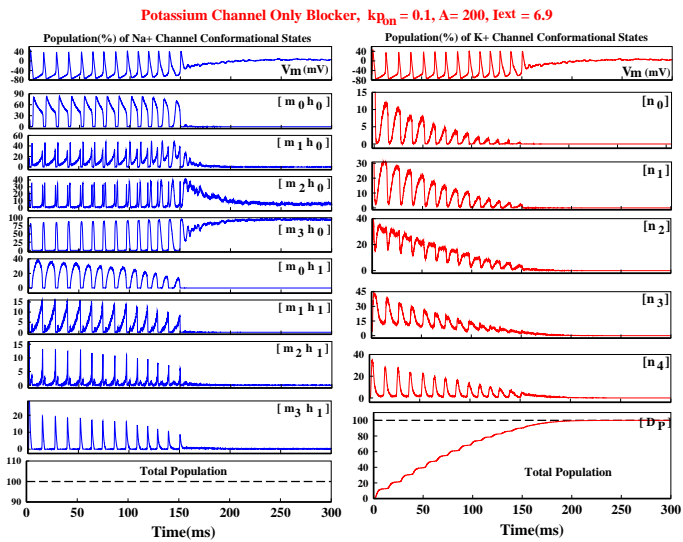}
\caption{\textbf{Gating dynamics in presence of potassium only blocker:} The action potential and the corresponding population(\%) of every sodium(left traces) and potassium(right traces)  channel conformational states are plotted for potassium only blocker with k$_{on}$= 0.1 at A= 200.0 $\mu m^2$ and $I_{ext}$= 6.9 $\mu A/cm^2$.} 
\label{f10}
\end{figure}

Next we have plotted every population(\%) of the sodium and potassium channel states  in presence of potassium only blockers in left and right traces of figure (\ref{f10}) respectively for $kp_{on}=0.1$. It is very astonishing to see that the trends of population dynamics has been drastically changed from that of the sodium only blockers in figure (\ref{f9}). They show different occupancy dynamics. It clearly says that different types of blocking agents can considerably change the gating dynamics of the system. This inference is consistent with the earlier literature\cite{Wang, Strichartz}. It is a very important realization which could be obtained only from a Markov model oriented studies like this.  Another interesting feature is that the population of the drug bound state, $D_P$ very rapidly grows towards 100\%  for $k_{on}$=0.1, unlike the sodium only blocking. The sodium channel population is mostly trapped in m$_3$h$_0$ state, which is a closed state. The altered gating dynamics and hampered membrane potential originating from  disrupted ion conduction in presence of blockers make them trapped in this state, as already discussed in Section \ref{sec3}. This is a cross verification of our previous statement. 

% \newpage

\section{Our Kinetic Drug Blocking Model Versus Other Studies In The Literature}

Hodgkin-Huxley squid axon model is a very basic representation of a neuronal cell. Since 1952 many Markov  models with  more coupled states or more complexities have come up that mimic the biological responses more accurately. But due to the diverse nature of living cell types there exist different types of models for different systems. In this paper our goal is to understand kinetically how the  ion channel blockers affect a neuronal cell, in general. Besides it is simply not possible to study all kinds of model and compare them. As we are looking at two criteria mainly: the variety of firing rate dynamics that can be reproduced and the shape of action potentials affected by ion channel blockers, the Hodgkin-Huxley(HH) model is a well accepted model\cite{Brenton-Jamasb-Nataliya-Roman-Xinmeng} in this regard compared to threshold models such as the Leaky Integrate and Fire(LIF)\cite{Indiveri} or the Izhikevich model(IZH) \cite{Izhikevich}, Morris-Lecar model\cite{morris}, FitzHuge-Nugamo(FHN)\cite{FHN} because  the parameters of the HH model have biophysical significance. Moreover one can modify the original model parameters to easily include the cell response of different systems\cite{David-Willms}. The stochastic drug binding approach on HH model is a very good choice since it is capable of bringing out the essential physiological features of ion channel blocking phenomena which corresponds to many experimental observations till date tested on different types of systems. In the following we present the references of such experiments done. 

\begin{enumerate}
\item{\textbf{Broadening of APD:} From the study of  potassium only blockers and the local anesthetic blocking that have more blocking potency to potassium channel than sodium blockers(case of $kp_{on}>ks_{on}$), we have seen that these blockers tend to broaden or delay the action potential duration. There exist lots of experimental evidences that support this finding. TEA, 4-AP(4-Aminopyridine), CTX (charybdotoxin) etc mainly known for potassium channel blocking potency show broadening of action potential in  demyelinated rat sciatic nerve\cite{Kocsis}, rat  hippocampal pyramidal cell\cite{Storm}, rat superior cervical sympathetic neurons\cite{Brown1}, on supraoptic neurons\cite{Hlubek}, on  Hippocampal CA3 Neurons\cite{bean1} and mammalian central neurons\cite{bean2}. The effect of Dendrotoxin, a potassium blocker from mamba snakes, on cerebellar basket cells, has also shown a delayed action potential duration\cite{Kullmann}.}

\item{\textbf{Multiple spike discharge:} It is also verified that 4-AP can lead to multiple spike discharge, spontaneous impulse activity and alteration of refractory periods\cite{Kocsis}. These results are totally consistent with our frequency response curves in figure (\ref{f6}) and action potential plots in presence of potassium blockers. From the figure \ref{f4}(b) and (c) we can also see that the spike train shows sudden spontaneous and enhanced spiking density temporally(spike train seen to be very dense) as also seen in ERG-K$^+$ channel blockade\cite{Gullo} and ABP(Ankyrin-binding peptide) blockade on Neural KCNQ(Kv7) channels\cite{Brown}. The effect of channel blockers on effective ionic current loop is also consistent with literature\cite{bean1}.}

\item{\textbf{Shortening of APD:} On the other hand  the shortening of action potential duration in presence of sodium channel blockers is more pronounced for cardiac cells, as far as our literature survey is concerned. Shortening of the action potential by ion channel blockers has been observed in several systems like sheep cardiac purkinje fibers\cite{Saikawa}, in rabbit cardiac Purkinje fibers\cite{Colatsky}, in dog ventricular cardiomyocytes\cite{Nanasi}, in guinea pig ventricular myocytes\cite{Belardinelli} etc. Antiarrhythmic agents such as Lidocaine, Phenytoin, Mexiletine, Tocainide etc  sodium channel blockers shortens the action potential duration\cite{mcgrawhill-perez}. Antiarrhythmic agents  such as bretylium, amiodarone, ibutilide, sotalol etc predominantly block the potassium channels, thereby prolonging repolarization\cite{mcgrawhill-perez}.  The mechanism of cardiac cells are quiet different due to the presence of other important types of ion channels and also the shape and length of action potential is quite different than neuronal cell . We do not compare our simple Hodgkin-Huxley action potential result to the cardiac action potentials. Hodgkin-Huxley model shows good agreement with the neuronal cells. We just mention that cardiac cells also show similar effects as the neuronal cells show in presence of channel blockers.}

\item{\textbf{Altered gating dynamics:} The  state  transitions  that underlay  the  gating  process(opening, closing or inactivation etc)  are  altered  by  local  anesthetics\cite{Wang}  and that  such altered gating itself  becomes the  essential  modulator  of local  anesthetic  block\cite{Strichartz}. From the results discussed for gating dynamics in figure (\ref{f9}) and (\ref{f10}), we have also shown that the sodium and potassium blockers have very distinct population dynamics. Thus both type of drugs drastically change the gating dynamics.}
\end{enumerate}

\subsection{Comparison Between Langevin and Markov Model Simulation}

The transient properties such as spiking interval, coefficient of variation of spiking due to channel blockers has been studied using Langevin description of stochasticity\cite{hangii3} in HH model. Here we have provided a brief comparative study between the original deterministic description of HH model using gating variables(m,h,n without white noise terms added), Langevin description(equation \ref{fox}) used in reference \cite{hangii3}  and our Gillespie simulation of HH-Markov chain model used in this paper. Ignoring the channel noise fluctuation we want to see how close the Langevin\cite{hangii3} and Gillespie simulation we adapted matches each other with or without the presence of blockers.  In Langevin description the dynamics of the gating variables are considered to be stochastic as follows,
\begin{equation}
\dot{z}= \alpha_z(V)(1-z)-\beta_z(V)z+ \eta_z(t), 
\label{fox}
\end{equation}
where $z=n,m,h$. $n$ represents the potassium and $m$ and $h$ together represents sodium gating dynamics as considered by Hodghkin-Huxley originally.  $\eta_z(t)$(s) are independent Gaussian white noise which makes the sodium and potassium gating dynamics stochastic in nature. The strength of the individual noises are given as follows\cite{hangii3}:
\begin{equation}
<\eta_{z}(t) \eta_{z}(t^\prime)>= \frac{2}{N_{i}}\frac{\alpha_z(V)\beta_z(V)}{[\alpha_z(V)+\beta_z(V)]}\delta(t-t^\prime),
\end{equation}
where $i= N_{Na}$ for z=m and h and $i= N_{K}$ for z=n. The blocking of sodium and potassium channel conductance are considered as fractional conductance\cite{hangii3} as follows,
\begin{equation}
G_K(V,t)=  g_K^{max}x_K n^4 \hspace{0.5cm}\text{and}\hspace{0.5cm} G_{Na}(V,t)=  g_{Na}^{max}x_{Na}m^3h,
\label{drug}
\end{equation}
where the factors $x_K$ and $x_{Na}$ are the fractions of working or unblocked ion channels
to the overall number of potassium and sodium channels, respectively. $0 < x_{K/ Na} \le 1$. Solving the conductance equation (\ref{drug}) and equation (\ref{mem}) one obtains the action potential using this Langevin type of description. 

\begin{figure}[h!]
\centering
\includegraphics[trim={0.001cm 0 0 0},clip,width=9.5cm,angle=0]{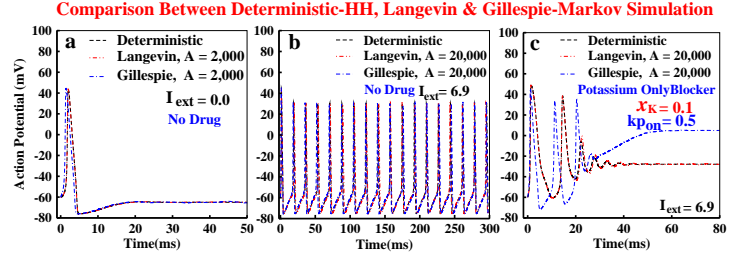}
\caption{\textbf{Comparison between original Deterministic-HH, Stochastic Langevin and Markov-Gillespie drug binding schemes: channel noise ignored:} In figure (a) the action potential generated from Deterministic-HH, Langevin and Markovian simulations has been compared with out the presence of drug and external current and channel noise. In figure (b) similar plot has been done in presence of an external current. In figure (c) we have compared three results in presence of very high potassium channel blocking affinity.} 
\label{f11}
\end{figure}

\textbf{Similarity:} From figure \ref{f11}(a) it is seen that without the presence of drug and stimulus both Langevin and Markov description matches the deterministic-HH description when channel number fluctuation is ignored with high patch size which also validates that our simulation-code results to be correct. Next in figure (b) we have shown that the Gillespie and Langevin model even in presence of stimulus, matches each other when the channel noise is ignored. However as the external current is applied the determiistic match requires higher patch size, such as A= 20,000 $\mu m^2$.  Next, in figure (c) we have compared the models in presence of high drug binding affinity(for example, we showed the potassium blocking case here only). In high patch size where the channel noise is ignored it is expected that the Langevin approach should produce similar result to that of the model we considered having high drug binding affinity. Now as seen from figure(c), we find that for obvious reason the original deterministic-HH description matches with Langevin description exactly showing that the action potentials, at this very high blockade region($x_{K}$=0.1), quickly damps down and as the refactorization is hampered no further spikes could regenerate. Now this result is compared to the very high affinity region of our model with kp$_{on}$=0.5. We can see that the Langevin simulation shows similar trend of action potential termination as our Markov model does. Although they do not match each other exactly, not expected to match each other either(blocking schemes are very different) but their nature is mostly same. Thus our model reduces to the Langevin model to an extent. The difference in the termination potential arises due to their inherent model differences. The termination potential in Langevin scheme is around -30 mV which is also unsuitable for further action potential generation.  

\textbf{Difference:} The difference between the Langevin scheme and our Markov scheme of drug binding is that the Langevin scheme discards a certain fraction of the total number of sodium or potassium channel and their contribution to the total conduction from the patch at the very beginning. The action potentials are then calculated on the basis of remaining number of  available channels, thereby assuming that during the time of observation, the number of blocked channels do not change, or as if the drug binding kinetics is in a standby mode. Actually in real situation during the course of  observation the available number of open channels or number of blocked channels change due to the drug binding. With higher affinity the channels  are blocked in a faster rate and vice verse.  This is the main reason why the Langevin approach can not show the gradual change in shape or APD with time. As the Langevin approach does not have any specific states of the channel(closed, open or inactivated state etc) a lot of information regarding the population dynamics, altered gating dynamics is lost which makes our Markov model approach a better alternative for studying kinetic drug blocking model. Most importantly, the intricate details of various drug binding schemes such as closed or inactivated state blocking etc can also be  adapted and implemented using simple consideration of suitable drug-bound states which are not possible in the Langevin approach. Also the Gillespie simulation of Markov model gives the opportunity to calculate the population of individual states which opens the door of carrying out various nonequilibrium thermodynamical analysis\cite{kp2018, hong} such as entropy production rates, free energy change etc of different binding schemes also. In short our Markov model based study has some advantages over Langevin approach  in the context of state specific ion channel blockade.

%\newpage
\section{Conclusion}
The definite presentation of specific channel states in the Markov model helps to describe not only the macroscopic current contributing for the action potential, but also the probability and transitions of each channel state, which gives a mechanistic and detailed link between the whole-cell action potential and the structure or function of ion channels. Thus Markovian simulation is an essential part of modern day model based elcetrophysiological and pharmaceutical investigations. Here we have tried to understand how one can theoretically bring out the essential features of ion channel blocking.  By the suitable adaption of Gillespie algorithm along with the direct numerical simulation of voltage dynamics we have been able to show how a simple extension of Hodgkin-Huxley model can prove to be an important tool to understand the effect of ion channel blockers to predict the important drug binding features like altered spiking frequency, altered duration of action potential of a neuronal cell and altered gating dynamics etc. This study establishes a link between the theoretical understanding of drug binding kinetics and the observed experimental findings\cite{Storm, Brown1, Hlubek, bean1, bean2, Kullmann, Gullo, Brown, Saikawa, Colatsky, Nanasi, Belardinelli, mcgrawhill-perez}.  Our study gives an opportunity to further investigate different types of drug binding features that can originate from other types of drugs such as closed state blockers or inactive-state blockers.  The major conclusions of this study in kinetic drug binding scenario  is as follows. 

\begin{enumerate}

\item{In the case of  only sodium channel blocking although the spiking activity decreases with blocking affinity, the train of action potentials generated are of complete shape because there are plenty of active potassium channels available to bring the system back to its resting state. However, in case of potassium blocking at higher value of  binding affinity, due to the lack of active potassium channels the shape of action potential at the end can not complete. Consequently  the re-factorization process gets hampered which eventually affects the sodium channel activation process to  regenerate  more spiking activity. In case of sodium channel blockers spiking activity slowly decreases mainly due to absence of adequate number of available open sodium channels, where as in case of potassium channel blockers the repolarization leading to re-factorization process plays a vital role in destroying spike generation process.  }

\item{We have also shown two distinct types of spiking trends in presence of two types of blockers. For sodium blocker we observe that the spiking frequency falls off  exponentially with the increasing binding affinity. In contrast,  with  increasing potassium blocking affinity, initially  the spiking frequency increases towards a maximum and then it gradually falls off exponentially. This initial increase in spiking activity arises due to the dominance of increased internal potassium channel number fluctuations or channel noise over total ionic conductances. In case of local anesthetics or dual blockers where the drugs have equal binding affinity to both  the  channels, show multiple maxima in spiking frequency trend. A drug with more binding affinity to a particular channel shows similar spiking frequency trends as in the case of only sodium or potassium blockers. However there exists a critical value of affinity for potassium channels above which irrespective of mechanisms all spiking activity is destroyed due to incomplete re-factorization process.}

\item{Both the sodium and potassium blocking agents considerably changes the gating and population dynamics of the system which we could show using our description.}

\item{Our approach satisfactorily agrees as well as provides suitable explanation regarding the experimentally observed change in action potential duration in presence of channel blockers. Sodium channel blockers shorten the action potential duration and the potassium blockers delays it. However in case of dual type blockers we have shown that potassium blockers play influential role by prolonging the repolarization process ultimately.}

\end{enumerate}

Thus using our basic mechanisms of ion channel blocking  one can develop a systematic understanding of the physiological effects of channel blockers for simple neuronal cell with sufficient details  which opens the possibility of exploring many other important drug binding features with the incorporation of desired level of structural and functional details. %Also this study once again proves the beauty of the gold standard Hodgkin-Huxley model. 

%\pagebreak

%\section*{Reference}
%\begin{thebibliography}{99}
%\clearpage

\section*{Appendix}
\subsection*{A1: Gillespie Algorithm for Site Selective Ion Channel Blocking}
To consider site selective binding of sodium or potassium blockers in a single neuron the Gillespie's  algorithm we used to study the drug binding kinetics has been implemented using the following steps.\bigskip \\

1. Initially we have fixed the number of sodium and potassium channels to be simulated. $N_{Na}=\rho_{Na} A$ and $N_K=\rho_K A$ are the numbers of sodium and potassium ion channels in a particular patch of area A $\mu m^2$ with sodium channel density, $\rho_{Na}$= 60 $\mu m^{-2}$ and potassium channel density, $\rho_{K}$= 18 $\mu m^{-2}$\cite{hangii2-hangii1-hangii4} .  

2. Then at a resting membrane potential, i.e. at -70 mV the steady state values of the gating variables n,m and h are solved using the following steady state equations\cite{hodg},
\begin{equation}
n= \frac{\alpha_n}{\alpha_n+\beta_n}, m= \frac{\alpha_m}{\alpha_m+\beta_m} \hspace{0.2cm}\text{and}\hspace{0.2cm}
h= \frac{\alpha_h}{\alpha_h+\beta_h}.
\end{equation}

3. Next at t=0 we have assigned the population of the drug-bound state, $D_{Na/K}=0$ for respective sodium and potassium blocker simulations. All the other 13 states are binomially distributed\cite{Skaugen},
$$ N^{kj}_{Na}=\binom 3j \hspace{0.1cm}h^k (1-h)^{(1-k)} m^j (1-m)^{(3-j)} N_{Na}, $$
\begin{equation}
N_K^j=\binom 4j \hspace{0.1cm} n^j(1-n)^{(4-j)}N_K,
\end{equation}
where $N^{kj}_{Na}$ and $N_K^j$ denotes the population of the states where k number of h gates, j number of m gates for sodium channel and j number of n gates for potassium channels are open. Here $N^{kj}_{Na}$ and $N_K^j$ are integers which fluctuate around their expected values.

4. Then we put an initial voltage of -60 mV and start the simulation for a paticular value of $I_{ext}$. First we calculate the individual propensities, $a_{K}(n_j \rightarrow n_{j'})$ for potassium ion channel and $a_{Na}(m_jh_k\rightarrow m_{j'}h_{k'})$ for sodium channels of all the 30 reactions (including the drug-bound state). Here ($m_jh_k \rightarrow m_{h'}k_{j'}$) or $(n_j \rightarrow n_{j'})$ indicates the transition between neighboring states. The propensities $a_{Na}(m_jh_k\rightarrow m_{j'}h_{k'})$ or $a_{K}(n_j \rightarrow n_{j'})$ for individual transitions are expressed as the transition rate multiplied by the population of the state from which the transition is taking place.  As for example, the transition from the state $m_1h_0$ to $m_2h_0$ has the propensity, $a_{Na}(m_1h_0\rightarrow m_2h_0)=2 \alpha_m N_{m_1h_0}$. Similarly for potassium channel the propensity for the transition between $n_3$ to $n_4$ is given by, $a_K(n_3 \rightarrow n_4)=2\alpha_n N_{n_3}$. Only the backward transitions from $D_{Na}$ or $D_{K}$,  the propensities has been given a constant value as mentioned earlier. This way we calculate 30 propensities corresponding to 30 reactions.

5. Next we calculate the total propensity, $a_{T}$ by summing over all the  propensities as calculated in the previous step. The sum can be expressed simply for sodium channel blockers as, $$a_{T}=\left[\sum_{\nu=1}^{22} a_{Na}^{\nu}\right] +\left[\sum_{\nu=1}^8 a_{K}^{\nu}\right].$$ 
and for potassium channel blockers,
$$a_{T}=\left[\sum_{\nu=1}^{20} a_{Na}^{\nu}\right] +\left[\sum_{{\nu}=1}^{10} a_K^{{\nu}} \right].$$ 

6. Next we calculate the random time required for the next transition to occur by calculating $\tau$, given as 
$$ \tau=\frac{1}{a_{T}}\ln(1/r_1),$$
where $r_1$ is a pseudo-random number called from an  uniform distribution[0,1]. As soon as the $\tau$ is obtained, the time is incremented by $t=t+\tau$.

7. In the next step we calculate which one of the 30 reactions has taken place in that $\tau$ time. An integer $\mu$ is assigned which designates the transition number. So $\mu$ varies from 1 to 30 in our case including the drug-binding step. Then  another random number, $r_2$ from an uniform distribution[0,1] of unit interval is called. Then we calculate the quantity, $(r_2 \times a_T)$. Then the transition number $\mu$ is calculated using the following relation $$\sum_{\nu=1}^{\mu-1}a^{\nu}_{Na/K} < (r_2 \times a_T)\leq \sum_{\nu=1}^{\mu} a^{\nu}_{Na/K}.$$ This is actually adding the successive propensities, such as $a_{Na}(m_0h_0\rightarrow m_1h_0)+ a_{Na}(m_1h_0\rightarrow m_0h_0)+ a_{Na}(m_1h_0\rightarrow m_2h_0)+ a_{Na}(m_2h_0\rightarrow m_1h_0)+.....+ a_{K}(n_0\rightarrow n_1)+a_{K}(n_1\rightarrow n_0)+ a_{K}(n_1\rightarrow n_2)+ a_{K}(n_2\rightarrow n_1)+....$ under the $\mu$ do-loop ($ \mu=$ 1 to 30) until their sum is equal or just exceed ($r_2 \times a_T$), and the loop number index or the transition index, $\mu$ is then set equal to the loop index or transition number index of the last $a_{\nu}$ term added. This is how the reaction taken place is identified. 

8. As soon as the reaction number index, $\mu$ is identified, the population of the states associated with that reaction is updated by $\pm 1$. The population of the  state from which the transition occurred is updated with -1 and the population of the state where to the transition has occurred is updated with +1.

9. Then the sodium and potassium conductances are calculated using equation \ref{cond} and the corresponding sodium,$I_{Na}$ and potassium , $I_K$ ionic currents are given by the equation
\begin{equation}
I_{K}=G_K(V,t) (V-E_K) \hspace{0.5cm}\text{and} \hspace{0.5cm} I _{Na}=G_{Na}(t)(V-E_{Na}).
\label{ion-curr}
\end{equation}

10. Then membrane potential is simply integrated with the calculated time step 
$\tau$ as follows,
$$V_i=V_{i-1}+\frac{1}{C_m}\left[(I_{ext}-I_{int})\tau\right],$$
where $I_{int}=I_{K}+I_{Na}+I_{L}$. %We have taken here the $I_{ext}=6.9$.

11. Then again the process is repeated from step 4, with the new value of membrane potential and updated transition rates and propensities at the obtained value of membrane potential, V. 

12. To calculate the spike frequency of action potentials we run the program for a very long time and then convert the spike number in Hz unit. %The program is written in Fortran 90.  

Using the above mentioned algorithm  we have written a  Fortran 90 code and complied in GFORTRAN to study the physiological effect of drug binding on action potentials.  \\ %The Fortran 90 code is provided as a supplemental material named ``\textbf{code.f90}". 

\subsection*{A2: Time Dependent Propensity vs. Constant Propensity}
 Here we want to mention that the method that we have adapted for the calculation of the propensities is a very popular and widely used method and often claimed to be an exact method for simulating action potenial using Gillespie algorithm. However contrary to this popular belief this method is not an exact method. In systems where the rate constants or the transition rates are time dependent through voltage change[$k_{z}(V(t)), k=\alpha \hspace{0.1cm}\text{or}\hspace{0.1cm} \beta, z=m,n,h$] or  change with time due to temperature change or volume change, propensity functions does not remain constant between reactions, i.e. $a^{\nu}_{Na/K}(t)=a^{\nu}_{Na/K}(X(t),t)$, where $X(t)=\lbrace [m_0h_0](t),..,[m_3h_1](t),[n_0](t),..,[n_4](t)\rbrace$ is the population state vector of different states\cite{anderson2007}. When the propensity functions depend only on the state of the system i.e. $a^\nu_{Na/K}(t)=a^\nu_{Na/K}(X(t))$ the Gillespie algorithm calculates the time until the next transition takes place by considering the first transition time of $R_\nu$(total number of transitions) \textbf{time homogeneous} transitions. However, as the transition rates become time dependent, the first firing time has to be calculated from  $R_\nu$ \textbf{time-inhomogeneous} transitions. The amount of time that must pass until the next transition takes place, $\tau$, is given by the distribution function\cite{anderson2007}:
\begin{equation}
1-\exp{\bigg(-\sum_{\nu=1}^{R_{\nu}}\int_t^{t+\tau} a_{Na/K}^\nu(X(t),s)ds\bigg)},
\end{equation}       
where $X(t)$ is constant in the above integrals as no reactions take place within the time interval $[t, t+\tau)$. Using the above equation, $\tau$ is obtained by first letting $r_1$ be uniform(0,1) and then solving the following equation:
\begin{equation}
\sum_{\nu=1}^{R_{\nu}}\int_t^{t_+\tau}a_{Na/K}^\nu(X(t),s)ds=\ln(1/r_1).
\label{hard}
\end{equation}
Then the transition that occurs at that time is chosen according to the probabilities\cite{anderson2007} $a_{Na/K}^{\nu}(X(t), t+\tau)/a_T$, where $a_T=\sum_{\nu=1}^{R_{\nu}} a_{Na/K}^{\nu}(X(t), t+\tau)$. So it is seen that this time dependent case is very different from the time homogeneous case which are most frequently used in literature where $\tau$ is exponentially distributed as $P(\tau)=a_T(X(t)\exp\big[-{a_T(X(t))\tau}\big]$, where $a_T(X(t))=\sum_{\nu=1}^{R_\nu} a^\nu_{Na/K}(X(t))$ and the probability that the next reaction is $\nu$-th reaction is given as ${a_{Na/K}^{\nu}(X(t))}/{a_T(X(t))}$.  However, solving equation(\ref{hard}) by both analytically and numerically is extremely hard and time consuming. Owing to the difficulty of solving equation (\ref{hard}) one can bypass it by using next reaction method or modified next reaction method\cite{anderson2007}. However those methods are also restricted to very few number of channels. The rigorous Gillespie algorithm remains a challenge for the stochastic Hodgkin-Huxley model. For very low patch size where very few number of  channels are considered the Gillespie algorithm with constant propensity assumption is very questionable, indeed it is wrong. However, for many channels with very large total propensity of a large population of channels the typical $\tau$ becomes so small that one can consider propensities to be approximately constant. The larger is the number of ion channels the better the approximation works. The reason of this brief discussion is to just inform the readers that the widely claimed exact one is basically an approximate method, not exact method. It diverges from the true action potential trajectories as time propagates(as seen using Morris-Lecar model in reference \cite{anderson2015})  for sufficiently small number of channels. Although it is not at all bad for a patch size of A= 200$\mu m^2$. We have used the constant propensity method. The results with this much of patch area  is negligibly affected and moreover our aim is to understand how the physiological effects of drug blocking can be obtained in general. Theoretical development on exact simulation is not our focus here. For better understanding about the discussion between the piece wise constant propensity and time dependent propensity readers are advised to read the following three references: \cite{anderson2007,anderson2015,anderson2011}.

\section*{Acknowledgment}

K. Pal wants to thank the anonymous referees for their valuable suggestions and The Department of Biotechnology, Govt. of India for fellowship.

%\clearpage
%{\footnotesize \bibliography{reference.bib}}
%\bibliographystyle{apalike}
%\end{document}
\newpage
%\section{Supplemental Material}
%%%%%%%%%%%%%%%%%%%%%%%%%%%%%%%%%%%%%%%%%%%%%%%%%%%%%%%%%%%%%%%%%%%%%%%%%%

\pagenumbering{arabic} 
\setcounter{page}{1}
\renewcommand\thefigure{\arabic{figure}}  
\setcounter{figure}{0} 
\renewcommand{\thefigure}{S\arabic{figure}}
\begin{center}
\textbf{\Huge{Supplemental Material}}
\end{center}

%\section*{Supplemental Material}

\textbf{1. Effective ionic current}: The effective ionic current is given as, I$_{int}=I_{Na}+I_K+I_{L}$.
\begin{figure}[h!]
\centering
\includegraphics[trim={0.001cm 0 0 0},clip,width=9.0cm,angle=0]{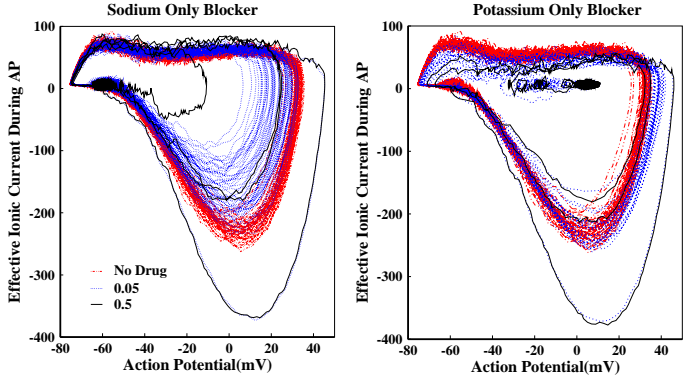}
\caption{ Effective ionic current is plotted here in presence of sodium and potassium only blockers for k$_{on}$=  0.05 and 0.5 has been plotted with the case when no drug is present. Here A= 200.0 $\mu m^2$ and $I_{ext}$= 6.9 $\mu A/cm^2$. The left panel is for sodium only blockers and the right panel is for potassium only blockers. }
\label{S1}
\end{figure}

%%%%%%%%%%%%%%%%%%%%%%%%%%%%%%%%%%%%%%%%%%%%%%%%%%%%
%newpage
\textbf{2. Dual Blockers, case: $ks_{on}=kp_{on}$}. The trend of action potential termination for the case, $ks_{on}=kp_{on}$ is shown below. 
\begin{figure}[h]
\centering
\includegraphics[trim={0.001cm 0 0 0},clip,width=9.0cm,angle=0]{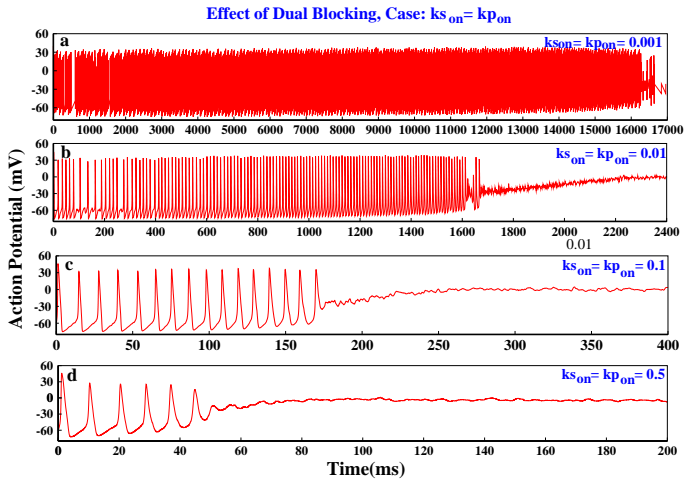}
\caption{The action potential for the case $ks_{on}=kp_{on}$ is plotted for affinity, $ks_{on}=kp_{on}$= 0.001 to 0.5 with A= 200.0 $\mu m^2$ and $I_{ext}$= 6.9 $\mu A/cm^2$.} 
\label{S2}
\end{figure}
%%%%%%%%%%%%%%%%%%%%%%%%%%%%%%%%%%%%%%%%%%%%%%%%%%%%%%%%%

\newpage
\textbf{3. Dual Blockers, case: $ks_{on}>kp_{on}$} The action potential for the case $ks_{on}>kp_{on}$ shows very noisy action potential trains. The peak height seems to fall gradually. Thus the spikes which have peaks above -10 mV has been considered in the paper works.
\begin{figure}[h]
\centering
\includegraphics[trim={0.001cm 0 0 0},clip,width=9.0cm,angle=0]{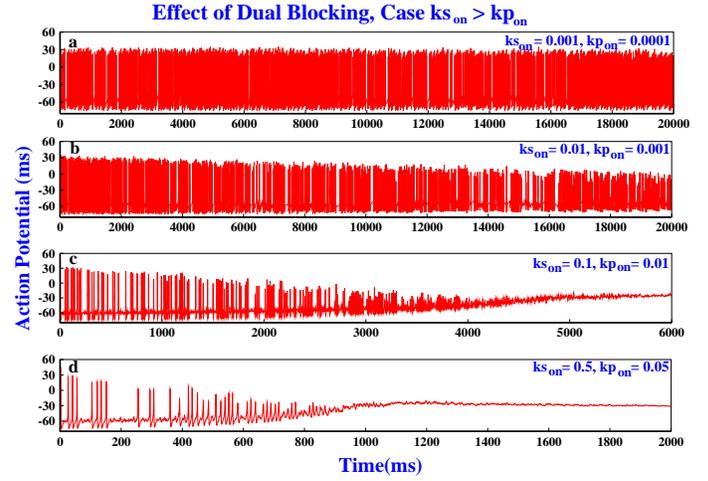}
\caption{The trend of action potential termination the case, $ks_{on}>kp_{on}$ is plotted here with A= 200.0 $\mu m^2$ and $I_{ext}$= 6.9 $\mu A/cm^2$.} 
\label{S3}
\end{figure}
%%%%%%%%%%%%%%%%%%%%%%%%%%%%%%%%%%%%%%%%%%%%%%%%%%%%%%%%%

\textbf{4. Dual Blockers,  case: $kp_{on}>ks_{on}$} is similar to that of potassium only blockers.

\begin{figure}[h!]
\centering
\includegraphics[trim={0.001cm 0 0 0},clip,width=9.0cm,angle=0]{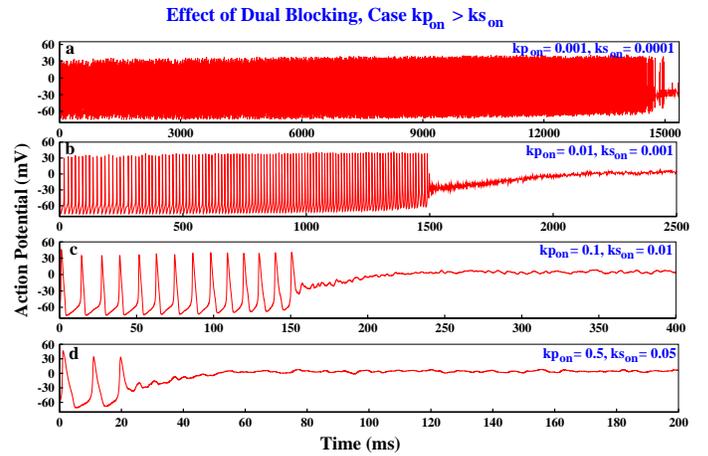}
\caption{The action potential for the case $kp_{on}>ks_{on}$ is plotted here for various affinities with A= 200.0 $\mu m^2$ and $I_{ext}$= 6.9 $\mu A/cm^2$.} 
\label{S4}
\end{figure}

%%%%%%%%%%%%%%%%%%%%%%%%%%%%%%%%%%%%%%%%%%%%%%%%%%%%%%%%%

\end{document}